\documentclass[11pt]{article}

\newcommand\dtmi{\ensuremath{\Delta_T \textrm{I}}\xspace}

\usepackage{PRIMEarxiv}

\usepackage[utf8]{inputenc} 
\usepackage[T1]{fontenc}    
\usepackage{hyperref}       
\usepackage{url}            
\usepackage{booktabs}       
\usepackage{amsfonts}       
\usepackage{nicefrac}       
\usepackage{microtype}      
\usepackage{lipsum}
\usepackage{fancyhdr}       
\usepackage{graphicx}       

\usepackage{enumerate}
\usepackage{amsmath}
\usepackage{makecell}
\usepackage{xspace}
\usepackage{tabularx}

\title{Using Monte Carlo Tree Search to Calculate Mutual Information in
High Dimensions
}

\author{
  Nick Carrara \\
  Physics Department\\
  University of California, Davis\\
  Davis, CA 95616\\
  \texttt{nmcarrara@ucdavis.edu} \\
   \And
  Jesse Ernst \\
  Physics Department\\
  University at Albany, SUNY\\
  Albany, NY 12222\\
  \texttt{jae@albany.edu} \\
}

\begin{document}
\maketitle


%
\begin{abstract}

Mutual information is an important measure of the dependence among
variables.  It has become widely used in statistics, machine learning,
biology, etc.  However, the standard techniques for estimating it often
perform poorly in higher dimensions or with noisy variables.  Here we
significantly improve one of the standard methods for calculating mutual
information by combining it with a modified Monte Carlo Tree Search.  We
present results which show that our method gives accurate results where
the standard methods fail. We also describe the software implementation
of our method and give details on the publicly-available code.

\end{abstract}

\begin{section}{Introduction~\label{sec:intro}}
Mutual information (MI)~\cite{coverthomas} is widely used to quantify
the strength of the relationship among variables.  In a simple case, if
a set of elements is characterized by two variables $V_1$ and $V_2$,
then it is the amount by which knowing the values of $V_1$ for the
objects reduces the entropy in the values of $V_2$. Informally, MI
measures how much the values of $V_1$ reveal about the values of $V_2$.
As expected, it is symmetric: $I[V_1;V_2]=I[V_2;V_1]$.

MI is more broadly useful than the correlation coefficient
\cite{coverthomas, Gencaga2014survey.mi.vs.correlation} because while
the correlation coefficient assumes a linear relationship between the
variables, MI measures the strength of the relationship without any
assumption about its underlying functional form.  MI is particularly
useful within machine learning (ML) applications as one typically has a
large number of variables with complex relationships among them.
Further, one generally has samples from the parent distributions, not
the distributions themselves, and so the functional relationships among
the variables are unknowable.

Throughout this paper, we will focus primarily on a standard
\emph{binary classification problem}.  Here, one has elements
(\emph{events}) that belong to one of two classes (\emph{signal} and
\emph{background}), and for each event one has the values for a set of
variables (called features or discriminating variables) for estimating
the event's class. In practice, a set of features is often used as input
to a neural network or other classification algorithm to estimate an
event's class.

MI is important in judging the discriminating power of features because the MI between the event
class and the feature is the amount by which knowing the value of a
feature reduces the entropy of the class. Or, informally: How well can
the values of the feature predict the class?  Measuring the MI between
features and class values is often used to choose the most effective
features; a process known as feature
selection~\cite{Guyon2003feature.selection.intro,
Sindhwani2004feature.selection.mi, Hanchuan2005feature.select.mi,
Battiti1994mi.featureselect, Bonev2008mi.featureselection}. The most
common procedure is to simply rank features based on the MI between
their values and the class
values~\cite{linsker.1988.self.org.perceptual.network,Vergara2013feature.select.mi.review,mondal.2020.mi.using.minmax,Estevez2009feature.select}.
There are also various heuristics for judging groups of
variables~\cite{steeg.2016.information,mondal.2020.mi.using.minmax,
Sombut2012feature.selection.mi, Torkkola2003feature.extraction.mi,
Fleuret2004fast.feature.selection}, although these become intractable as
the number of dimensions grows.

In an earlier paper, we emphasized that the MI between all available
features and the class values (if one can measure it accurately), gives
a useful estimate of the upper limit of separability between the
classes~\cite{Carrara2017upperlimit}. That is, it quantifies how well
any algorithm that uses all of those features as input can ever perform
as a classifier. Equivalently, it can be viewed as quantifying how much,
if any, discriminating information from the features was lost by the
transformation that used them, a quantity which we call \dtmi.

While extremely useful for problems with a small number of variables, MI
is less useful as the number of variables grows because the methods for
estimating it rapidly become unreliable with increasing dimension.  And
unfortunately, that is the situation in a wide range of class-separation
problems \cite{whiteson2014deeplearning}. The most widely used method
for calculating MI was introduced by Kraskov et al.\
(KSG)~\cite{ksg2004mi}, but even with only a few dimensions, it can
significantly underestimate MI if variables are noisy and/or have
significant correlations among them.  It has been noted that in these
cases, KSG and similar techniques are more useful for identifying the
presence or absence of correlations than they are for actually finding
an accurate value of MI
\cite{Gao.2014.mi.strongly.correlated,gao.2015.mi.gaussian}.  This
greatly limits MI's usefulness for setting a limit on class
separability~\cite{Carrara2017upperlimit}, or for feature selection. In
this work, we describe a method we developed, based on KSG, that allows
us to accurately estimate MI in higher dimensions, in the presence of
noise, and in the presence of correlations among variables.

In high dimensions and in the presence of noise and/or strong dependence
among variables, KSG's estimate of MI is known to fall significantly
below the true
value~\cite{Shashank.2016.knn.distance.entropy.estimate,czyz.mi.benchmarks.2023}.
One could in principle remedy this by somehow choosing a subset of
variables, or transformations of them, such that each variable would
have independent information about the class values. The remaining
variables, which have no independent information about the event class,
would then be discarded. Note that if one included these discarded
variables in the calculation, the true value of MI would not change, but
KSG's estimate of MI would be falsely low. Alternatively, one could test
all subsets of variables to find the subset with the largest KSG
estimate of MI. Unfortunately, either of these brute-force approaches
would quickly become intractable for more than a few dimensions.

Our approach is to efficiently search for the subset of variables with
the largest returned MI value from KSG.  We do this by recasting the
problem of choosing included/excluded variables into the problem of
searching the tree of possible moves in turn-based strategy games. This
will allow us to use the Monte Carlo Tree Search method
(MCTS)~\cite{browne2012mcts.survey} which was developed specifically to
improve computers' ability to play these games by allowing them to
search a decision tree for the optimum move without doing a brute-force
evaluation of every possible path from the current state of the game to
its conclusion~\cite{silver2016go}.  Our approach therefore, is to: 1) recast the problem of
searching for the optimum set of variables into a problem that is
similar to turn-based games, and then 2) modify the MCTS method so that
it can quickly find the set of variables that maximizes KSG's estimate
of MI.

\subsection{Organization} In Section~\ref{sec:mi} we briefly review MI
and several related quantities.  We also discuss MI's connection to the
upper limit of separability between event classes and to the lower limit
on classification error rates.  We also discuss MI's importance in the
closely related problem of feature selection.  In Sections~\ref{sec:ksg}
and~\ref{sec:ksg_problems} we review the standard methods for estimating
MI, and the problems they have in high dimensions. In
Section~\ref{sec:mcts} we discuss how we use a modified MCTS method in
combination with KSG to estimate MI even in high dimensions and in the
presence of noisy variables and give some details on our software
implementation and on how to access the publicly-available code.  In
Section~\ref{sec:results} we compare the results of our method to those
of standard methods.  The conclusion (Section~\ref{sec:conclusion})
includes a glossary of the most used terms and acronyms,

\end{section}

\begin{section}{Mutual Information, Separability, and Classification
Errors}~\label{sec:mi}

Mutual information (MI) is a robust measure of the dependence among
variables.  Because it characterizes both linear and non-linear
dependencies among variables and is also invariant under smooth and
uniquely invertible transformations of variables, it has become a
centrally important data analysis tool in a wide range of fields from
machine learning, epidemiology, domain generation, deep learning,
biology,
etc.~\cite{Piras.mi.deep.learning.2023,Hjelm2018deep,Grabowski402750,li2022invariant}.

The definition of MI follows directly from the basic building block of
information theory: Shannon Entropy, or $H$~\cite{shannon}. For a random
discrete variable $x_i \in X$, 
\begin{equation}
    H[X]=-\sum_{i=1}^{|X|}p(x_i)\log p(x_i)\label{shannon_entropy}
\end{equation}
describes the uncertainty in $X$, or equivalently, the amount of
information per element that one would need to perfectly predict the
values of $X$.  If one has two variables, the related quantity
\begin{equation}
    H[X|Y]=-\sum_{j=1}^{|Y|}p(y_j)\sum_{i=1}^{|X|}p(x_i|y_j)\log
    p(x_i|y_j)\label{shannon_conditional}
\end{equation}
is the conditional entropy and is the amount of entropy, or uncertainty,
remaining in the values of $X$ after one is given the values of $Y$.
The mutual information $I[X;Y]$ of $X$ and $Y$,  is then defined in a
straightforward way as the amount by which the entropy of $Y$ is reduced
by knowing the values of $X$, or vice versa.  Thus,
\begin{equation}
    I[X;Y] = I[Y;X] = H[Y] - H[Y|X] = H[X] - H[X|Y]\label{MI_decomposition}
\end{equation}
and quantifies the amount that the values of $X$ tell you about the
values of $Y$. We use base-2 for logarithms throughout, which gives $H$
and $I$ the units of bits.

In a typical binary classification problem, where one uses $X$ to
estimate $Y$, these quantities are particularly simple.  For each event,
$X$ may be single-valued or multi-dimensional, but $Y$ has one of only
two values.  Thus, if the population of events is equally distributed
between the two classes, then $H[Y]=1$ and $I[X;Y] = 1 - H[Y|X]$.

For the binary classification problem, $I[X;Y]$ is equivalent to the
Jensen-Shannon divergence (JSD) between the $X$ distributions for the
two classes of events. Intuitively, a larger difference between the
$X$-distribution for signal and the $X$-distribution for background
corresponds to $X$ revealing more about the event class $Y$.

\begin{subsection}{The Upper Limit of Separability}~\label{subsec:seplimit}
In a previous paper~\cite{Carrara2017upperlimit}, we pointed out that if
one can accurately measure MI in a multidimensional space, then its
value and an appeal to the Data Processing Inequality
(DPI)~\cite{coverthomas} allow one to: 1) set a testable upper-limit on
the separability between classes of events in a binary classification
problem, and 2) more rapidly find a set of discriminating variables.

As a reminder, the Data Processing Inequality (DPI) formalizes the
intuitive notion that an algorithm using $X$ to predict $Y$ can not
reduce the entropy of $Y$ below some fundamental amount of information
that the values of $X$ hold about the values of $Y$. That fundamental
amount is the MI between $X$ and $Y$.  That the DPI is an inequality
rather than an equality reflects the fact that an unwise use of the
values of $X$ for predicting the values of $Y$ could discard
information.  An extreme example would be a function that casts all
values of $X$ into a single value, resulting in the loss of all
information that $X$ could have revealed about $Y$.  For smooth
invertible functions of $X$, the equality will hold.

Combining an accurate measurement of MI with the DPI allows one to set
an upper-limit on class separability and then use it to objectively
check the effectiveness of any discriminating algorithm.  In a standard
binary classification problem, one has a vector of discriminating
variables $x \in X$ for each event, and then uses an ML algorithm to combine
them into a single value $f(x)$ which estimates the class value $y \in Y$.
Using these to objectively test an algorithm is straightforward: 1)
Compute $I[X;Y]$, which we refer to as I\textsubscript{initial} or $I_i$.  $I_i$ is
the information that the vectors of discriminating variables $X$ can
reveal about the class values $Y$. 2) Use the $X$ values in a
discriminating algorithm to produce output values $f(X)$ that predict
the event classes $Y$. 3) Compute $I[f(X);Y]$, which we refer to as
I\textsubscript{final} or $I_f$. $I_f$ is the information that the algorithm output
values $f(X)$ can reveal about the class values $Y$. According to the
DPI, $I[f(X);Y] \leq I[X;Y]$ (i.e., $I_f \leq I_i$) where the equality holds
only if the algorithm is optimal and has not discarded any information
that $X$ could have revealed about $Y$. $I_i - I_f$, which we define as
\dtmi, is the information loss caused by the transform.  For a
classification model, it is the average number of bits of information
per event that the model has lost.

\dtmi is an objective way to judge whether a classification algorithm is
optimum.  This approach is very different from the usual practice of
assuming that a classification algorithm is optimum when neither
additional training nor alternate algorithms seem to improve a figure of
merit. Less than optimal results ($\dtmi>0$) can occur when, for
example, an algorithm settles into a local minimum, when it has too few
free parameters to fully model the space of the input vectors $x$, or
when the algorithm is poorly suited to the problem.

An accurate value of $I_i$ is also valuable on its own as a metric for
choosing discriminating variables (feature selection)
\cite{Vergara2013feature.select.mi.review,doquire.mi.feature.selection.2012,bennasar.mi.feature.selection.2015}.
Feature selection has numerous heuristic procedures involving adding or
dropping variables and then, after each change: 1) retraining the
algorithm, 2) reprocessing the events through the algorithm, and 3)
measuring the change in some figure of merit. These ad-hoc approaches
are needed because a brute-force comparison of all possible combinations
of potential discriminating variables is generally intractable.  There
is no perfect non-brute-force solution to this problem, but if one can
quickly and accurately compute $I_i$, then one can rapidly compare sets
of potential features without ever training an ML model on them.  As we
discuss below, we have been able to compute $I_i$ roughly two orders of
magnitude faster than training and testing an ML model, and hence
potential sets of features can be evaluated extremely quickly.

Note that comparing $I[X;Y]$ to $I[f(X);Y]$ where $Y$ is discrete, as
described above, differs from the Infomax
strategy~\cite{Linsker.1988.infomax,bell.infomax.1995}, where one uses
$I[X;f(X)]$, where both $X$ and $f(X)$ are continuous distributions.
With Infomax, as $f(x)$ is developed, one uses the differential
(continuous) entropy $I[X;f(X)]$ as a figure of merit to minimize the
amount of information lost as $X$ is transformed into $f(x)$.  Rather
than following Infomax, we take advantage of the fact that our problem
has discrete answers.  This lets us compare the discrete Shannon
entropies $I[X;Y]$ and $I[f(X);Y]$.  That comparison allows us to
quantify absolutely the effectiveness of a classifier, and quantify
absolutely the usefulness of potential discriminating features.

The challenge in using $I_i$ for feature selection or \dtmi for
quantitatively judging a model, is in accurately measuring $I_i$. While
calculating $I_f$ is simple because both the algorithm output $f(X)$ and
the class values $Y$ are generally one dimensional, calculating $I_i$ is
much more difficult because the dimension of the $X$ vectors is equal to
the number of discriminating variables, which is often large.  And all
current standard methods for computing MI struggle in high dimensions.
Thus much of the rest of this paper will focus on accurately computing
$I_i$.

\end{subsection}

\begin{subsection}{Figures of Merit}~\label{subsec:fom}
Before continuing with the calculation of $I_i$, it is worth briefly
discussing MI as a figure of merit (FOM).  MI is defined clearly as the
amount by which the entropy of one variable is reduced by knowing the
value of another variable (or set of variables).  In the discussion
above, we call an algorithm optimum if $\dtmi=0$, and thus its output
has as many bits of information about the class label as the original
variables had.  And we compare two potential sets of discriminating
variables by comparing how many bits of information they hold about the
class label.  But for a particular problem, other FOM's may be
preferred~\cite{zhao2013beyond}. Examples include: the largest expected
statistical significance, the smallest error rate, the smallest
false-positive rate within some constraint on the false-negative rate,
etc. These will all tend to improve with improving MI, but it isn't
generally possible to directly convert from one FOM to another.

The relationship between error rate and MI has been studied extensively.
For a given $I[X;Y]$, the lower bound and upper bound on the error rate
for $Y$ are given by Fano's limit \cite{coverthomas, scarlett2019fano}
and the Hellman-Raviv limit \cite{hellman.raviv1970bound} respectively.
Fano's limit states that for a given MI, no algorithm can be found with
an error rate lower than $P_e > H^{-1}(1-I[X;Y])$, where $H^{-1}$ is the
inverse of the Shannon entropy. The Hellman-Raviv limit states that it
is always possible to find an algorithm with an error rate better than
$H(1-I[X;Y])/2$.  These are shown in figure~\ref{fig:fano_raviv}.
\begin{figure}[!htb]
\centering
\includegraphics[width=3.6in]{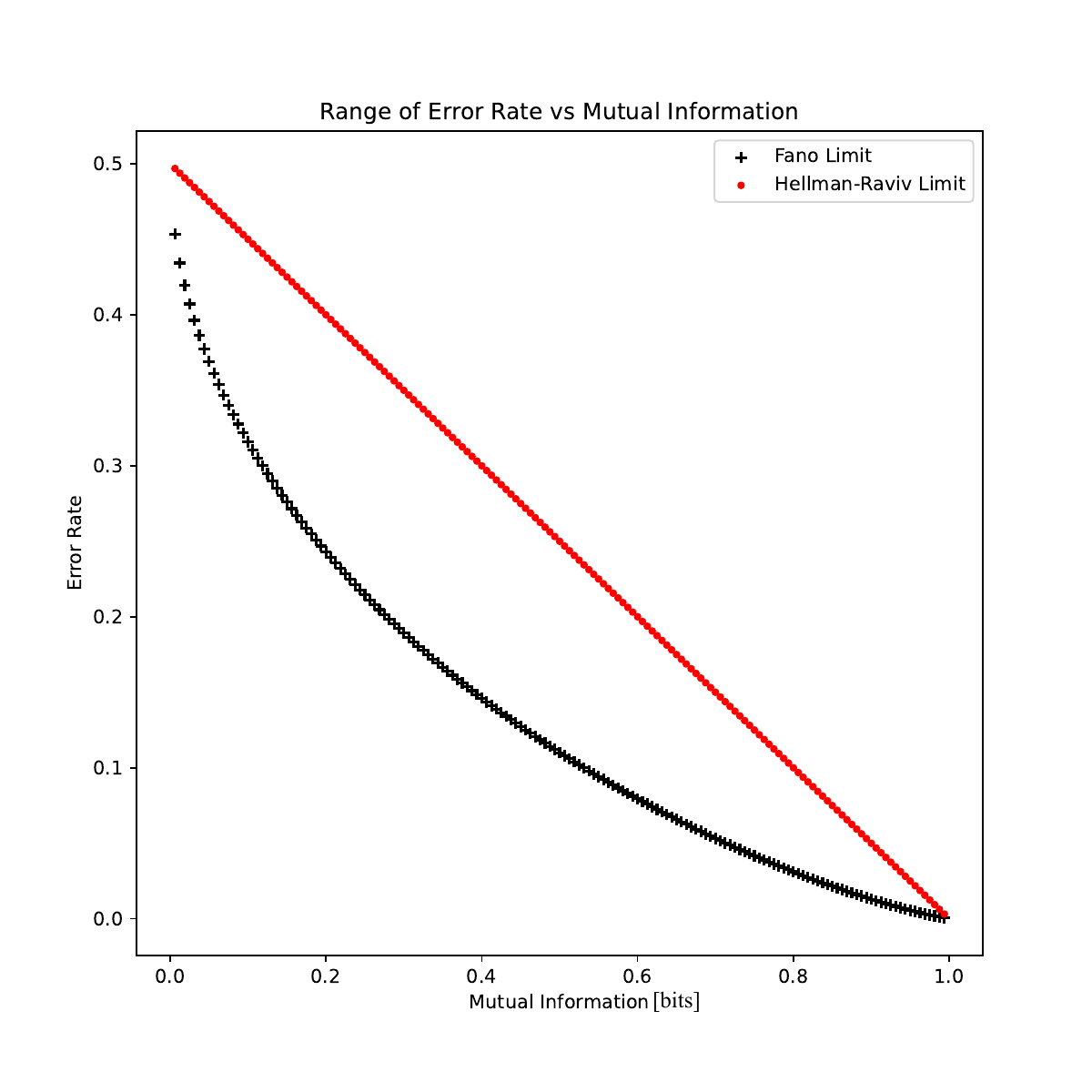}
\hfill
  \caption{Fano and Hellman-Raviv limits on error-rate versus MI in bits
  for a binary classifier. The two curves define the lower and upper
  bounds respectively on the error rate, $P_e$, as a function of MI for
  any classifier, i.e. $H(1 - I[X;Y])/2 > P_e > H^{-1}(1 -
  I[X;Y])$.~\label{fig:fano_raviv}}
\end{figure}

Note that both limiting cases make sense.  In the limit where MI=0 and
thus $X$ gives no information about $Y$, all algorithms will have a 50\%
error rate. In the limit of MI=1, $X$ gives complete information about
$Y$ and it is always possible to find an algorithm with a 0\% error
rate. Between these two cases, there is a range of possible error-rates
for a given MI.

Using \dtmi to decide whether an algorithm is optimum remains valid,
independent of one's choice of FOM. An algorithm that discards
information ($\dtmi>0$) will have a lower (or at least not better) FOM
than one with ($\dtmi=0$) for any reasonable FOM.  Using $I_i$ for
feature selection is more subtle.  Using $I_i$ guarantees that the
output from an optimum algorithm will have the largest possible MI with
the class label, but figure~\ref{fig:fano_raviv} shows that there will
be a range of corresponding error rates.  So in principle, it might be
possible to choose a set of variables with lower $I_i$ but which
nonetheless has a smaller error rate.  We suspect that this will be
uncommon, but see no way to prove it.  It likely depends on details of
the distributions, and we don't consider it further
\end{subsection}

\end{section}

\begin{section}{MI Estimators\label{sec:ksg}}

While widely useful, MI is notoriously difficult to compute.  The
 fundamental problem is that one generally doesn't have closed
 functional forms for $p(x|y)$, $p(x)$, or $p(y)$ but rather has finite
 i.i.d.\ samples drawn from them $z_i = (x_i, y_i) \in \mathcal{Z}$, $i
 = 1,\dots,N$, requiring the use of numerical methods for integration.
 The calligraphic symbols $\mathcal{X} \subset X$ are used to denote
 sample distributions (point sets) drawn from the underlying space $X$.

 \subsection{Histogram Methods}
 The simplest approach for computing MI from i.i.d.\ samples is to
 create a histogram for the three sample distributions $\mathcal{X}$,
 $\mathcal{Y}$, and $\mathcal{Z} = \mathcal{X}\times \mathcal{Y}$ by
 choosing some suitable binning such that each space is partitioned into
 $h_{\mathcal{X}}$, $h_{\mathcal{Y}}$ and
 $h_{\mathcal{Z}}=h_{\mathcal{X}}h_{\mathcal{Y}}$ bins.  The value of
 the probability in each bin is simply given by the number of points
 which land in that bin relative to the others, i.e. $P_x(i) \approx
 n_x(i)/N$, $P_y(j) \approx n_y(j)/N$ and $P_z(i,j) \approx n_z(i,j)/N$
 where $i = 1,\dots,h_{\mathcal{X}}$ and $j = 1,\dots,h_{\mathcal{Y}}$.
 The MI is approximated by the sum,
 \begin{equation}
     I[\mathcal{X};\mathcal{Y}] \approx \sum_{i,j}P_z(i,j)\log
     \frac{P_z(i,j)}{P_x(i)P_y(j)}.
 \end{equation}
 While the histogram method converges to the true MI in the limit of
 $N\rightarrow \infty$ and bin sizes going to zero, it becomes
 intractable as the dimension of the spaces increases.

 As discussed by KSG, methods for computing MI based on cumulant
 expansions or entropy maximization \cite{roberts2001, hyvarinen2001},
 have poor accuracy.  Methods based on kernel density estimators (KDE)
 can do well in limited circumstances, but often perform no better than a
 linear correlation coefficient~\footnote{See Section IV.A in
 \cite{ksg2004mi}}.

\subsection{Non-parametric Entropy Estimators}
The non-parametric estimators described here are a family of $k$-nearest
neighbor ($k$NN) methods, which estimate the local density at each point
in $z_i = (x_i,y_i)$, $x_i \in \mathcal{X} \subset X$ and $y_i \in
\mathcal{Y} \subset Y$ by using information from its $k$-nearest
neighbors.  The first goal is to construct an unbiased estimator of the
Shannon entropy $H[X]$, which has the form
\begin{equation}
    \hat{H}[X] = -\frac{1}{N}\sum_{i=1}^N\log
    p(x_i),\label{unbiased_estimator}
\end{equation} where $p(x_i)$ is the \textit{true} density at $x_i$ so
that $\hat{H}$ converges to the \textit{true} entropy as $N\rightarrow
\infty$\footnote{It is apparently provably impossible to find an
unbiased method \cite{Holmes2019Esimation,Paninski2003mi.estimate}}.

\subsubsection{KL Estimator} The KSG algorithm for estimating MI is
based on an approach developed by L.F.\ Kozachenko and N.N.\ Leonenko
(KL)for estimating the Shannon entropy $H[X]$ of a continuous variable
$x \in X$~\cite{KL}.  We will reproduce the derivation here, which
begins by writing the probability that the $k$th-nearest neighbor of the
point $x_i \in \mathcal{X}$ exists in a small spherical shell of radius
$\varepsilon/2$ and thickness $d\varepsilon$ so that there are $k-1$
points lying within the shell at points $r_i < \varepsilon/2$ and that
there are $N - k - 1$ points lying at distances $r_i >
\varepsilon/2 + d\varepsilon$.  This distribution is given by the
multinomial,
\begin{align} P_{\varepsilon}(x_i)d\varepsilon =
    &\frac{N!}{1!(k-1)!(N-k-1)!}\nonumber\\ &\times p_i^{k-1}(1 -
    p_i)^{N-k-1}\frac{dp_i}{d\varepsilon}d\varepsilon\label{multinomial},
\end{align} where $p_i$ is the probability mass integrated over the
inner spherical region,
\begin{equation} p_i = \int_{r_i < \varepsilon/2} dx\, p(x),
\end{equation} and where $p(x)$ is the \textit{true} probability
density.  The distribution in (\ref{multinomial}) has the interesting
property,
\begin{equation}
    \langle \log p_i \rangle = \int d\varepsilon\,
    P_{\varepsilon}(x_i)\log p_i = \psi(k) -
    \psi(N),\label{expected_log}
\end{equation} where $\psi$ is the digamma function ($\psi(x) =
\Gamma(x)^{-1}d\Gamma(x)/dx$)\footnote{When $x$ is discrete, the digamma
function follows the recursion formula $\psi(x+1) = \psi(x) + 1/x$ and
$\psi(1) = -C$ where $C = 0.5772156\dots$ is the Euler-Mascheroni
constant.  We can then easily compute $\psi(N) =
\sum_{n=1}^{N-1}\frac{1}{n} - C$ for any integers $N$.}, and $N$ is the
number of points in the data set.

\paragraph{KL density assumption ---}The main assumption for all
estimators comes from the next step, in which one assumes some form of
the local density $p(x_i)$ in order to approximate $p_i$ for each point.
For the KL-estimator, and later the KSG estimator we assume that the
local density around each point $x_i$ is uniform (i.e. the density
$p(x_i)$ does not vary much in the $\varepsilon/2$ ball),
\begin{equation} p_i \approx
    c_d\varepsilon^dp(x_i),\label{uniform_assumption}
\end{equation} where $c_d$ is the volume of the unit ball in
$d$-dimensions and $\varepsilon$ is the distance to the $k$-neighbor.
As we will see in the next section, it is more convenient to use
rectangular volumes which are aligned with the coordinate axes of $X$
and $Y$ when defining the volumes of the $p_i$'s.  In this coordinate
system, the form of eq.~(\ref{uniform_assumption}) changes slightly to
\begin{equation} p_i \approx
    \left(\prod_{m=1}^d\varepsilon^i_{m}\right)p(x_i),\label{uniform_box}
\end{equation} where $\varepsilon^i_m$ is some distance along the
$m$th-dimension for the $i$th point.  Using this form, we have from eq.~(\ref{expected_log})
\begin{align}
    \psi(k) - \psi(N) = \sum_{m=1}^d\langle \log
    \varepsilon^{i}_m\rangle + \langle \log p(x_i) \rangle.
\end{align} Inserting the above into the unbiased estimator
(\ref{unbiased_estimator}), one arrives at the KL entropy estimator,
which contains a single parameter $k$,
\begin{equation}
    \hat{H}_{KL}^k[X] = -\psi(k) + \psi(N) +
    \frac{1}{N}\sum_{i=1}^N\sum_{m=1}^d\log(\varepsilon^{i}_{m}),\label{KL_estimator}
\end{equation}

\subsubsection{KSG Estimator} To extend the KL algorithm to estimate MI,
one could naively plug in eq.~(\ref{KL_estimator}) to the MI
decomposition in eq.~(\ref{MI_decomposition}), which amounts to
estimating the entropy of $\hat{H}[X]$, $\hat{H}[Y]$ and
$\hat{H}[X\times Y]$ and combining them to give the MI,
\begin{equation}
    \hat{I}_{\mathrm{naive}}[X;Y] = \hat{H}[X] + \hat{H}[Y] -
    \hat{H}[X\times Y].\label{Inaive}
\end{equation}  However, as KSG point out in motivating their estimator,
this can result in a bias from applying the KL estimator with the same
parameter $k$ to spaces of different dimension and scale, which leads to
non-cancellation of the errors in the three entropy
estimates~\footnote{See Section I page 2 of \cite{ksg2004mi} for more
discussion.}.  One can see this by first applying the KL estimator to
the joint space $X\times Y$,
\begin{align}
    \hat{H}_{KL}[X\times Y] = &-\psi(k_{Z}) + \psi(N)\nonumber\\ &+
    \frac{1}{N}\sum_{i=1}^N\sum_{m_Z=1}^{d_X +
    d_Y}\log(\varepsilon^i_{m_{Z}}),\label{KL_joint}
\end{align} where $Z = X \times Y$.  Plugging eq.~(\ref{KL_joint}) and
the corresponding marginal estimates into eq.~(\ref{Inaive}) we find,
\begin{align}
    \hat{I}_{\mathrm{naive}}[X;Y] &= \psi(k_{Z}) + \psi(N) -
    \frac{1}{N}\sum_{i=1}^N\sum_{m_Z=1}^{d_X +
    d_Y}\log(\varepsilon^i_{m_Z})\nonumber\\ &-\psi(k_X) +
    \frac{1}{N}\sum_{i=1}^N\sum_{m_X=1}^{d_X}\log(\varepsilon^i_{m_X})\nonumber\\
    &-\psi(k_Y) +
    \frac{1}{N}\sum_{i=1}^N\sum_{m_Y=1}^{d_Y}\log(\varepsilon^i_{m_Y}).
\end{align} The nearest neighbor factors $k_{Z}$, $k_X$ and $k_Y$ are,
in general, all different, and imposing $k_Z = k_X = k_Y$ can result in
the volume terms $\langle \log \varepsilon_i\rangle$ not canceling and
accumulating errors.  KSG solves this problem by imposing that the
volume measures used in the joint space $Z$ is the same used in the
marginal spaces $X$ and $Y$.

\paragraph{KSG marginal assumption ---}Instead of choosing the same $k$
for each space, they simply choose one $k = k_{X \times Y}$ for the
joint space and then force the projected $\varepsilon^i$ in each
marginal space to be the same as in the joint space, i.e.
\begin{equation}
    \sum_{m_Z=1}^{d_X + d_Y}\log(\varepsilon^i_{m_Z}) =
    \sum_{m_X=1}^{d_X}\log(\varepsilon^i_{m_X}) +
    \sum_{m_Y=1}^{d_Y}\log(\varepsilon^i_{m_Y}),\label{marginal_projection}
\end{equation} which simply amounts to adjusting the $\psi(k_X)$ and
$\psi(k_Y)$ for each projected $p_i$ in the marginal spaces.  As noted
by KSG, fixing $k$ in the joint space requires the use of the
$L^{\infty}$ norm when defining $p_i$, which results in all distances
being the same, i.e., $\varepsilon_m^i = \varepsilon^i$.

With the above adjustment, the functional form for the KSG estimator is
\begin{equation}
    \hat{I}_{KSG}[X;Y] = \psi(k) + \psi(N) - \langle \psi(n_X +
    1)\rangle - \langle \psi(n_Y + 1)\rangle,\label{KSG}
\end{equation} where $n_x$ is the number of points falling in the
projected $p_i$ and $\langle \psi(n_x + 1)\rangle$ is the mean over all
points.
\end{section}

\begin{section}{KSG: Limitations and Improvements\label{sec:ksg_problems}}
While the KSG estimator in eq.~(\ref{KSG}) works well in some
applications, it is well known to consistently underestimate MI as noise
and/or number of dimensions increases.  Czyz et
al.~\cite{czyz.mi.benchmarks.2023} have studied a wide range of MI
estimators applied to forty different datasets with varying noise levels
and numbers of dimensions.  Of the 40 tests, KSG is accurate in 17
tests, underestimates MI (often significantly) in 23 tests, and
overestimates MI in zero tests.  Similarly, when introducing their G-KSG
method, Marx and Fischer~\cite{marx.2022} show that over a wide range of
datasets and dimensions, that the standard KSG method is either accurate
or underestimates MI. That KSG consistently underestimates rather than
overestimates MI when noise variables are added or as the data become
sparse is central to our method of improving it.

There are several assumptions used in the KL/KSG estimators that lead to
common failure modes:
\begin{enumerate}\label{ksg.assumptions}
    \item \textbf{KSG marginal projections} - The central feature of the
     KSG estimator is that equation (\ref{marginal_projection}) is
    imposed when calculating the marginal entropies $\hat{H}[X]$ and
    $\hat{H}[Y]$.  This feature is spoiled by the use of the naive estimator
    $\hat{I}_{\mathrm{naive}}$ in equation (\ref{Inaive}).

    \item \textbf{Uniform local densities} - The single assumption of
    the KL estimator in approximating the probability mass in equation
    (\ref{uniform_assumption}) is that the probability density $p(x_i)$
    around the point $x_i \in \mathcal{X}$ is uniform. If this
    assumption is not satisfied, it can indirectly lead to the failure
    of the first assumption, since for arbitrary forms of the density
    estimate in equation (\ref{uniform_assumption}) it is not
    immediately clear how to impose the cancellation of the resulting
    marginal volume terms. This problem occurs for several estimators
    based on KSG~\cite{gao.2015.mi.gaussian,
    Gao.2014.mi.strongly.correlated} which we discuss in
    Section~\ref{sec:ksg_fixes}.

    \item \textbf{Joint space $L^{\infty}$ box} - KSG uses the
    $L^{\infty}$ box to define $\varepsilon_i$ and hence $p_i$ in the
    joint space in eq.~(\ref{marginal_projection}).  This is not
    necessarily ideal for regions where the density is not uniform, as
    has been argued by several
    authors~\cite{Gao.2014.mi.strongly.correlated}.  If this assumption
    is not satisfied, then, again, the first assumption will not be
    satisfied.

    \item \textbf{Fixed k-neighbors} - A design choice of the KSG
    estimator and the KL estimator is to use a single $k_Z$ value for
    all points $z_i \in \mathcal{Z}$ in the joint space.  When densities
    have large variations in a space it may no longer be appropriate to
    use a fixed $k$ value for all points. Unlike the other assumptions,
    choosing a dynamic $k$ value does not spoil other KSG features.

\end{enumerate}

\subsection{Failure Modes of KSG\label{sec:ksg_fail_modes}} Most failure modes
result from some aspect of the curse of dimensionality since, with any
numerical method, we will be subject to limitations that come from
sparseness.  This has been explored in the case of KSG, to some extent
by the original authors as well as others.  Gao et
al.~\cite{gao.2015.mi.gaussian} showed that the KSG estimator is
consistent, but that its bias is dimension dependent and is of the order
$\mathcal{O}(N^{-\frac{1}{d_x + d_y}})$.

In the original KSG paper~\cite{ksg2004mi}, they show several examples
comparing their estimator with closed functional forms of the MI for
special distributions, and in each case where the estimate is
meaningfully different from the true value, it is an underestimate (see
their figures 7, 10, 11, 12 and 14).  Regions where MI is overestimated,
which occurs when the number of samples is large, only gives a
disagreement of a few percent.

While the remaining failure modes outlined below have several examples
showing their effects, there are currently no quantitative explanations
linking the behavior to the assumptions in the KSG estimator.

\subsubsection{Redundant Variables} 
We define ``redundant'' variables as those which are dependent on other
variables, but which do not introduce any new information.  As a simple
example, consider the case of two variables, $X_1$ and $Y$, where there
exists correlations so that $p(x_1,y) = p(x_1)p(y|x_1)$ and $I[X_1;Y] >
0$ (i.e. $p(y|x_1) \neq p(y)$).  Then, consider that with $X_1$ we add
an additional variable $X_2 = f(X_1)$ which adds no new information
about $Y$, but is highly correlated to $X_1$, i.e.
\begin{align}
    p(x_1,y) \rightarrow p(x_1,x_2,y) &= p(x_1,x_2)p(y|x_1,x_2)\nonumber\\
    &= p(x_1,x_2)p(y|x_1).
\end{align}
Then, the MI will remain invariant,
\begin{equation}
    I[X_1;Y] \rightarrow I[X_1\times f(X_1);Y] = I[X_1;Y].
\end{equation}
The authors have found that most situations involving redundant
variables are handled well by KSG, as long as the number dimensions is
small.

\subsubsection{Noisy Variables} We define ``noisy'' variables as those
that do not contain any additional information when added to a problem
but are also independent of any other variables.  Consider the example
from the previous subsection except now where $X_2$ is independent of $X_1$, i.e.
\begin{align}
    p(x_1,y) \rightarrow p(x_1,x_2,y) &= p(x_1,x_2)p(y|x_1,x_2)\nonumber\\
    &= p(x_1)p(x_2)p(y|x_1).
\end{align}
Then, as it should be \cite{Carrara.2020.GlobalCorrelation}, the MI will
also be invariant\footnote{Note the difference to the redundant case in
which $p(x_1,x_2) = p(x_1)p(x_2|x_1)$.},
\begin{equation}
    I[X_1\times X_2;Y] = I[X_1;Y].
\end{equation}
Because of the limitations of numerical methods variables with weak
correlations may be difficult to distinguish from noisy variables.  As
with redundant variables, it is clear from numerous
examples~\cite{Carrara2019} that noisy variables cause KSG to
underestimate MI.

\subsubsection{Non-uniform PDF's} The KL estimator, and hence the KSG
estimator, rely on the assumption in equation~(\ref{uniform_assumption})
that the local density $p(x_i)$ inside the ball $p_i$ is uniform.  As
shown by several authors~\cite{Gao.2014.mi.strongly.correlated,
gao.2017} this assumption can lead to underestimation of MI in cases
where the density varies largely over regions of the space.  It is
especially evident in regions around the boundary of a space, where the
$L^{\infty}$ box can extend into regions where the probability density
goes to zero.  Although the authors present some interesting solutions,
we do not pursue them here.

\subsubsection{Discontinuous Transformations\label{sec:homeomorphism}}
If one uses KSG to measure the MI between a set of variables and the
event classes, and then applies a non-homeomorphic transformation to the
variables, KSG will fail to find the same value of MI.~\footnote{We are
unaware of any method for measuring MI that is stable under
non-homeomorphic transformations. In the most extreme case, the data
processing inequality tells us that a set of variables that contain
information about event classes retain that information, in principle,
after applying a cryptographic hash function to them. But there is
clearly no known method that can indicate even that the transformed
variables contain a non-zero amount of information about the event
classes.}

\subsection{Attempts to Address KSG Failure Modes\label{sec:ksg_fixes}}
Attempts to improve KSG fall into one of several categories:
\begin{itemize}
    \item \textbf{Local density corrections} - Some suggested
    improvements to KSG attempt to address the density assumption
    pitfalls (items `b' and `c') as outlined at the start of
    section~\ref{sec:ksg_problems}.  These adjustments are concerned
    with the assumptions in eqs.~(\ref{uniform_assumption}) and
    \ref{uniform_box} which are replaced by some other method.
    \item \textbf{Dynamic k-nearest neighbors} - This type of adjustment
    changes the fixed $k$ value with one that depends on the local
    environment around each point $z_i$.  This involves some method for
    determining the $k_i$ at each point.
\end{itemize}
As shown by Marx and Fischer \cite{marx.2022}, the spoiling of the
$L^{\infty}$ condition in eq.(\ref{uniform_box}) causes MI to be
overestimated, which is undesirable.  Some examples of the \textit{local
density correction} are discussed below in
Sections~\ref{sec:ksg_fixes_llnc} and \ref{sec:ksg_fixes_kde}. An
example of \textit{dynamic k-nearest neighbors} is discussed in
Section~\ref{sec:ksg_fixes_gksg} and an approach using neural networks
is discussed in Section~\ref{sec:ksg_fixes_mine}.

\subsubsection{Local Non-Uniform Correction (LNC)\label{sec:ksg_fixes_llnc}}
As an attempted correction to KSG’s problem with using the $L^{\infty}$
box, S. Gao et al.\ proposed the \textit{local non-uniform correction}
(LNC) technique  \cite{Gao.2014.mi.strongly.correlated}. This technique
adjusts the unbiased estimator for MI by replacing the $L^{\infty}$
volume in the joint space with a volume computed from a PCA analysis.
The basic idea is the following: Consider a point $x_i$ whose
$k$th-neighbor is $x_k$.  With the collection of $k + 1$ points
including $x_i$, $x_k$ and all points closer than $x_k$, construct the
correlation matrix $C_{ij}$ and find its eigenvectors.  By then
projecting each point along the maximal eigenvectors, we can find a PCA
bounding box which is rotated and skewed with respect to the
$L^{\infty}$ box. The assumption in this case is that the rotated PCA
box is a better representation of the region of uniform probability
around $x_i$.  Once each volume is found, the MI is given by
\begin{equation}
    \hat{I}_{LNC} = \hat{I}_{KSG} - \frac{1}{N}\sum_{i=1}^N\log\frac{\bar{V}_i}{V_i},
\end{equation}
where $\bar{V}_i$ is the PCA volume and $V_i$ is the $L^{\infty}$
volume. Such an estimator has shown to give vast improvement to the
naive KSG method, however current results are limited to two dimensional
problems. The reason for this is its inability to deal with redundant
information. To see this, consider a two-dimensional problem in which
the variables $X \times Y$ have some non-trivial correlations. If we add
to X a redundant copy, $X \rightarrow X \times f(X)$, then the true
value of MI will be unchanged. If one naively uses the LNC method,
however, one will find that the MI increases. This is because the
volumes $\bar{V}_i$ will generally decrease when computed in the
redundant scenario and hence the LNC correction term will generally
increase \cite{Carrara2019}.

\subsubsection{Kernel Density Estimators\label{sec:ksg_fixes_kde}}
Gao et al.\ proposed a kNN density estimator~\cite{gao.2017}, called the
\textit{Local Likelihood Density Estimator} (LLDE), which weights
neighbors using parameterized Gaussian kernels whose bandwidths are
determined by the nearest neighbors.  The motivation for this
construction was to correct the biases of estimators near boundary
points of the underlying manifold.  One can easily see that any naive
KDE methods will underestimate the density near boundaries, since large
parts of the sampling region can be outside the manifold.  While not
directly a modification to KSG, the authors showed improvements to
vanilla KDE methods in simple cases.  In the study by Marx and Fischer,
they showed that the LLDE estimator can lead to a dramatic
over-estimation of MI, which makes it undesirable for our purposes.

Lord et al.\ proposed a geometric kNN density
estimator~\cite{lord.2018}, called \textit{Geodesic-KNN} (G-KNN), which
uses ellipsoids fit with PCA to describe the local density rather than
the Gaussians of the LLDE method and the $L^{\infty}$ box of KSG.  While
they show some decent results for low-dimensional examples, the G-KNN
method also tends to overestimate MI, particularly in high-dimensional
cases as shown by Marx and Fischer \cite{marx.2022}.

\subsubsection{Geodesic-KSG (G-KSG)\label{sec:ksg_fixes_gksg}}
A. Marx and J. Fischer introduced an extension to KSG~\cite{marx.2022}
   which uses geodesic distances to determine the $k$-nearest neighbors
   to use for each point in the joint space $X
   \times Y$.  This $k$-adaptive G-KSG method shows improvement over
   standard KSG in high-dimensions and where there are a large number of
   noisy variables.  Their estimate (like KSG) typically underestimates
   MI.

The G-KSG method works by first performing a manifold learning task on
the data through a Geodesic Forest (GF) algorithm, which is then used to
compute local geodesic distances that are used in place of the standard
$L^1$ distances assumed by KSG.  The benefit of this approach is that it
does not alter the assumptions of the KSG estimator.

\subsubsection{Mutual Information Neural Estimation
(MINE)\label{sec:ksg_fixes_mine}} The MINE approach uses the DPI to
estimate MI by first training a neural network, $f:X\rightarrow Y$, to
learn a suitable lower dimensional representation of the input space.
In this way, the value of $I_i = I[X;Y]$ is found by first fitting the
function $f$, and then estimated as $I_f = I[f(X);Y]$.  While this
scheme can work in principle, it assumes that the fitting of the
function $f$ reaches an optimum, i.e., that the learning algorithm has
found the global maximum so that $I_i = I_f$.  However, since no
calculation of $I_i$ is performed it cannot be known whether the
function $f$ actually achieves the equality.
\end{section}

\begin{section}{Combining MCTS and KSG~\label{sec:mcts}}

\subsection{Overview~\label{sec:mcts_overview}}
Here we give an overview of our modification to the Monte Carlo Tree Search
(MCTS) method, and of how we combine it with the KSG estimator to measure MI in
higher dimensions and in the presence of noise.  This section gives only the
details that are needed to understand the rest of the paper. Additional
implementation details are given in Section~\ref{sec:mcts_details}.

\subsubsection{Introduction to Monte Carlo Tree Search}
The Monte Carlo Tree Search (MCTS) method is a widely used algorithm for
estimating the optimum next move in a game by judiciously searching a small
fraction of the game tree.  The timeline of its continual development over the
past $\sim30$ years is well-documented in Table~1
of~\cite{browne2012mcts.survey}. At any point in a game, a player has $c_1$
choices for their move, which results in their opponent having $c_2$ choices
for their move (see Figure~\ref{fig:mcts}). This continues down the tree until
the players reach one of the final states of the game, at which point a winner
is determined.  These final states are represented by leaf nodes at the bottom
of the tree, and each corresponds to a unique and fully specified game from the
starting point. In all but the simplest games, the size of the tree makes it
intractable to search exhaustively for an optimum next move.  To estimate the
optimum move, the MCTS method builds and searches only the most profitable
portion of the game tree by repeatedly generating and evaluating multiple paths
through the tree, choosing its current path based on the success rates of
previous paths.
\begin{figure}[!htb]
    \centering
    \includegraphics[width=3.5in]{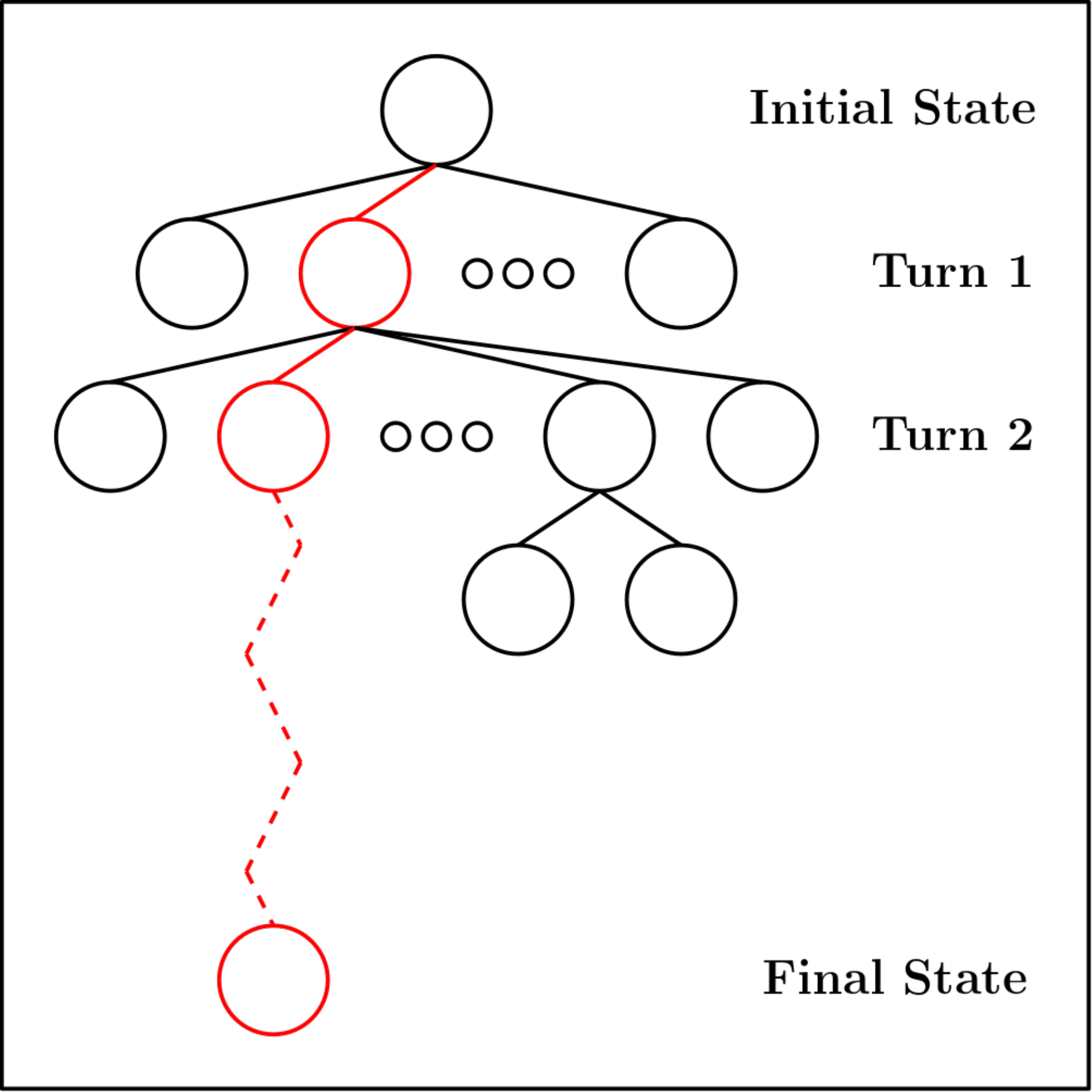}
    \caption{A search tree starts from an initial state, represented by the top
    node, and then branches according to the possible moves each player can
    choose.  Each path through the tree represents a unique sequence of moves
    from the starting point to the end of the game.}
    \label{fig:mcts}
\end{figure}

The process begins from the top node, which represents the current state
of the game.  In the first stage, called \emph{selection}, the forward
steps are chosen based on the \emph{tree policy} (see
Figure~\ref{fig:mcts}). Among the previously explored nodes, the tree
policy balances the benefit of stepping to a child node that has been
successful in previous paths against the possibility that a lower
performing node might, with more testing, ultimately be superior. The
tension between these priorities is known as exploitation vs.\
exploration and its study has been central to the development of
MCTS~\cite{browne2012mcts.survey,
silver2016go,kocsis2006bandit,reddy2016infomax.explore.exploit}.
Eventually the process will either reach the bottom of the tree
(unlikely in realistic problems) or will reach a node that has at least
one unexplored child node.  At this point, the path will step to one of
the unexplored child nodes. This stage is called \emph{expansion}.
Clearly all child nodes below this previously unexplored node will also
be unexplored. Following expansion is the \emph{simulation} stage, in
which the rest of the path to the end of the tree is chosen by the
\emph{default policy} (often simply randomly) and the winner is
recorded. This completion of the game using the default policy is called
a \emph{playout}. In the final stage, called \emph{backpropagation}, the
win/loss result becomes part of the statistics for all nodes that
participated in that path. Those statistics, through the policy
function, determine the next path through the tree. This process repeats
until a sufficiently strong next move is found or a budgeted computing
limit is reached. The MCTS method along with numerous variations is
summarized in more detail elsewhere~\cite{browne2012mcts.survey}.

Central to the MCTS method is the tension between exploitation and
exploration. There is an obvious benefit to further searching the
portions of the tree with higher win rates to more accurately judge the
best move.  At the same time, the opportunity cost of that exploitation
is that some other portion of the tree with potentially even better
performance may be left unexplored.  This idea appears in a broad range
of statistical decision making problems and is usually framed as a slot
machine with multiple arms (many-armed bandit).  In this problem, a
gambler has some number of trials in which to maximize their profit, and
each arm has some unknown profitability distribution.  The gambler wants
to continue to pull an arm that has so far been successful, but at the
same time wants to explore the possibility that some other less-tested
arm might ultimately be even more profitable. The optimum balance
between exploration and exploitation has been widely studied.  A paper
by Lai and Robbins~\cite{Lai.1985.bandit.regret} set a lower bound on
the rate at which \emph{regret}\footnote{The regret of a procedure is
the difference between its success and the success of an omniscient
player who only ever pulls the optimum arm.} grows with the number of
trials and thus created an absolute standard against which any decision
policy could be measured. MCTS methods incorporate these results in the
\emph{tree policy} function, which is used to choose the next step
forward among competing options.

\subsubsection{T-KSG} KSG could be used to accurately measure MI from a
dataset in high dimensions and with noisy variables if one had a way to
first remove the noisy variables from KSG's calculation. Because
increasing noise and increasing sparsity cause KSG to underestimate MI,
an equivalent goal is to find the subset of variables for which KSG
gives the largest estimate of MI.  Our approach (Tree-Search KSG, or
T-KSG) is to use MCTS to efficiently search the power set of the input
variables for the subset with the largest KSG-estimate of MI.

In our problem, we have $n$ discriminating variables (i.e., $x$ has
dimension $n$) and so have $2^n$ unique sets of variables.   We could in
principle use KSG to measure MI for each of the $2^n$ possible sets of
variables and use the largest value as our estimate of $I[X;Y]$. (We
would use the largest value from among the subsets because: 1) sets that
are missing variables with independent information about $y$ will have
their actual MI reduced by that amount of information, and 2) sets with
redundant and/or noisy variables will have a lower estimate of MI for
reasons discussed in Section~\ref{sec:ksg_fail_modes}.)

We can recast our problem as a decision tree by organizing the possible
variable sets as shown in Figure~\ref{fig:mist1}. Beginning at the root
node (level 1) with zero variables included in a set, one moves down the
tree and chooses to include (exclude) variable one by stepping left
(right). One continues stepping left (include) or right (exclude) at
each layer until one reaches the bottom of the tree.  Each leaf node
then corresponds to unique set of included variables. Rather than an
intractable search of the entire tree for the largest value of $I[X;Y]$,
we will use some of the ideas from the MCTS technique to search only a
portion of the tree.
\begin{figure}[!htb]
    \centering
    \includegraphics[width=3.6in]{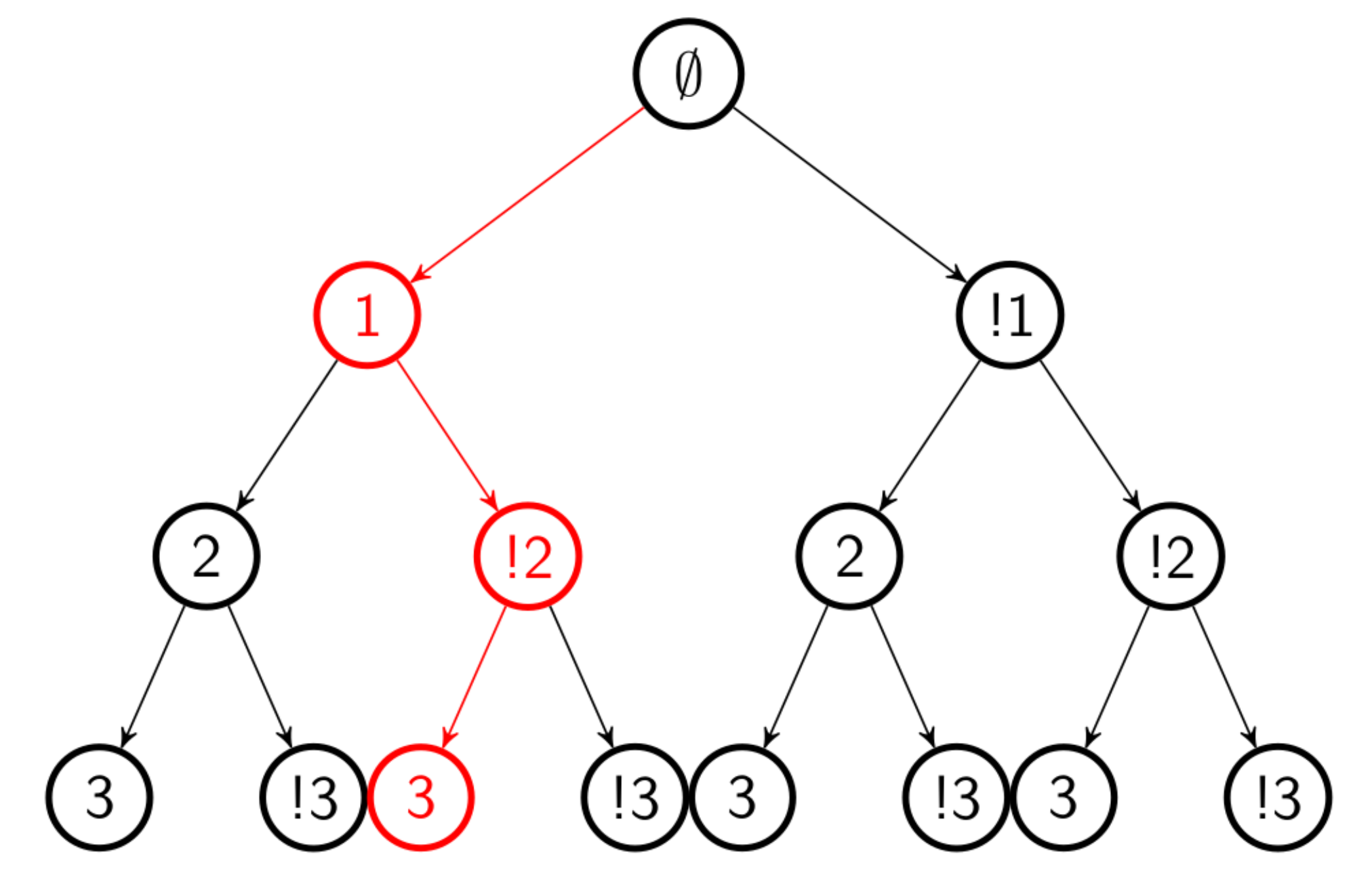}
    \caption{A search tree for organizing the possible subsets of three
    variables.  One begins with no included variables and at each step
    chooses either to include or to exclude the next variable.  Each
    path through the tree corresponds to a unique subset of variables.
    The red path, for example, corresponds to including variables 1 and
    3, and excluding variable 2. The Figure is from Chapter 6
    of~\cite{carrara.thesis.2021}.}
    \label{fig:mist1}
\end{figure}

The standard implementation of MCTS will not work for our problem. In
MCTS, each path has a binary result (win/loss), while we have a
continuous result (KSG's estimate of $I[X;Y]$). Further, the reward for
a path does not depend on the margin of victory, while we want to more
strongly favor paths with a larger value for KSG-estimated MI.   We
modify MCTS in two significant ways:
1) We make an MCTS variation that uses a continuous result rather than a
   discrete result, and 2) To speed optimization, we close off
   (\emph{gate}) sections of the search tree that perform poorly.

\emph{Discrete vs.\ Continuous:} Standard MCTS records a binary
(win/loss) result and back-propagates it up the tree, updating the
statistics of the nodes that were part of the path. In our case, when we
reach a leaf node, we use KSG to compute the MI of that path. That
continuous result then needs to propagate up the tree and become part of
the statistics for each included node. We thus need to: 1) choose the
figures of merit to be stored in each node, and 2) develop a policy
function for choosing futures paths based on those figures of merit.

\emph{Gates:} Noisy variables will cause KSG to underestimate MI, but the
amount of underestimation is unpredictable.  This can make convergence of the
tree search extremely slow, and so we developed a procedure to permanently
include/exclude variables based on whether the observed MI scores are
significantly higher when the variable is included/excluded in paths. More
detail on this is in Section~\ref{sec:mcts_details}.
Section~\ref{sec:mcts_details}.

\subsubsection{Limitations of T-KSG} Our T-KSG method has two primary
limitations. The first is a practical limit on the number of dimensions.
It can accurately compute MI in far more dimensions than KSG (see
Section~\ref{sec:results}), but it is still limited to hundreds, not
thousands of dimensions. We find that T-KSG is accurate at all numbers
of dimensions we have been able to test.  However, it can take nearly a
day to converge when dimensions reach $\sim 200$~\footnote{Tests used a
2019 Macbook with a 2.4 GHz 8-Core Intel Core i9 processor and 32 GB
2667 MHz DDR4 memory.\label{foot:specs}}.  And we expect that computing
time would continue to rise rapidly with increasing dimension, though we
haven't yet tried to estimate the computational complexity.

The second limitation is that transformations of variables must be
homeomorphic (smooth and uniquely invertible). As discussed in
section~\ref{sec:homeomorphism}, KSG's estimate of MI is only stable
under homeomorphic transformations of the variables.  And because T-KSG
uses KSG to evaluate MI for each candidate subset, it shares the
limitation that non-homeomorphic transformations of the variables will
change its estimate of MI.

\subsection{Implementation Details~\label{sec:mcts_details}}

Nothing in this section is needed to understand the sections that follow
it. Here we describe in more detail how our variant of MCTS works and
how it differs from standard MCTS. Our software implementation of T-KSG
allows for a wide range of variations on MCTS, but here we only describe
the specific variation and parameter settings that were used for the
results shown in Section~\ref{sec:results}. Full code, documentation,
and ready-to-run examples are available on
\href{https://github.com/jesse794/T-KSG}{GitHub}.

\subsubsection{Statistics Stored in Nodes}

In standard MCTS, each node stores the number of times it has been
included in a path and the percentage of those times that led to a
final-state win.  In our variation we need to store more detailed
information because for each path we are computing a continuous value
(MI) and using MCTS to maximize that value. First, recall, as shown in
Figure~\ref{fig:mist1}, that a node in the tree is a binary decision
point. The corresponding variable will either be included (step left),
or excluded (step right), and after a full path has been taken, KSG is
called to evaluate MI. Then, that MI value is propagated back up the
tree~\footnote{Throughout this section, when we say ``MI'', we mean
KSG's estimate of ``MI''.}. In each node, we store: 1) the number of
times the node's variable was included/excluded at that point in the
tree~\footnote{Note that the number of times a variable was
included/excluded by a specific node is not the same as the number of
times the variable was included/excluded in any path. More concretely,
the number of times variable 5 was included when variable 3 was excluded
is not the same as the total number of times variable 5 was ever
included in any path.}, 2) the mean MI and RMS values of MI when the
node's variable was included/excluded, and 3) the number of times that
the MI of the path was within some fixed percentage of the highest
recorded MI.

\subsubsection{Tree Policy, Default Policy, and Gate Policy}

\begin{itemize}
\item\textbf{Tree Policy} - In standard MCTS, the tree policy uses a
multi-arm bandit function to choose the next step forward in a path
based on the win/loss statistics of a node.  In plain terms, for our
tree with two options at each node, the tree policy creates competition
between the include side and the exclude side of a node.  If the node's
variable has independent information about the event class then paths
that follow its include side will generally have higher MI, while if it
is only noise, its exclude side will generally have higher MI, because
KSG will underestimate MI when noise is included.  As in standard MCTS,
our modified multi-arm bandit function generally favors the
higher-scoring side, but increasing weight is given to the side with
fewer visits as the number of visits become more imbalanced.  As with
all multi-arm bandit functions, this is intended to optimize exploration
vs.\ exploitation.  In our variation, when the search arrives at a node,
it chooses include vs.\ exclude by choosing the larger of $S_{inc} =
P_{inc} + 4\sqrt{\log(n_{both}/n_{inc})}$ and $S_{exc} = P_{exc} +
4\sqrt{\log(n_{both}/n_{exc})}$, where $P_{inc}$ ($P_{exc}$)is the
percentage of paths through the include (exclude) side of the node that
yield an MI within 5\% of the maximum found MI, $n_{both}$ is the total
number of paths that have passed through either side of the node, and
$n_{inc}$ ($n_{exc}$) is the total number of paths that have passed
through the include (exclude) side of the node.

\item\textbf{Default Policy} - In standard MCTS, the default policy is
applied when the search path reaches a node where none of its children
has been explored.  That of course also means that no children below
that point have been explored.  Here we can follow one of the standard
approaches where one simply chooses randomly among the options until one
reaches a leaf node at the bottom of the tree.  In our case, that
corresponds to choosing randomly whether to include/exclude all
remaining variables.  Then, as usual, that path is scored and its result
propagated into the statistics of the nodes that were in the path.

\item\textbf{Gate Policy} - In standard MCTS, all nodes in principle
remain accessible to future paths.  This causes significant problems for
T-KSG.  The issue is that when one includes noisy variables, the amount
by which KSG will underestimate MI can vary widely.  Unless one
gradually removes variables that lower KSG's estimate, converging to the
highest estimate can be extremely slow.

To handle this problem, we introduce gates that permanently include
(exclude) a variable that increases (decreases) MI more than one would
expect from fluctuations. There is no equivalent to this in standard
MCTS.  Roughly, we permanently include a variable by first finding the
percentage of paths that include the variable and have MI estimates that
are in the top 10\% of all paths. If a variable were neutral, one would
of course expect to find 10\% of its paths in the top 10\%. If  we find
that it is more than 10\% with confidence level of $3.5\sigma$, then the
variable is permanently included in future paths. Using the same
standard, if paths that exclude the variable have higher MI estimates,
then the variable will be permanently excluded.

\end{itemize}

\subsubsection{Other Related MCTS Modifications}
There are numerous MCTS variations, and hundreds of applications reported in
the literature.  We note two efforts that are somewhat similar to our work.

Gaudel and Sebag~\cite{gaudel.2010.mcts.feature.selection} study feature
selection as a reinforcement learning problem, where the optimum feature
set is the one with the smallest generalization error based on a reward
with low computing cost.  They additionally use MCTS to retain the
features in the most often visited path of a feature-selection search
tree. In contrast, our work focuses on using MCTS to calculate MI. An
advantage of our approach is that calculating MI for a set of variables
is extremely fast. As we discuss in section~\ref{sec:mi}, knowing MI
then lets us determine how well an ML algorithm could perform without
ever actually creating one. Further, T-KSG will allow us to select
approximately the smallest set of features that have this maximal MI
value.

Chaudhry and Lee~~\cite{Chaudhry.2018.feature.selection.mcts} study
feature selection using, as we do, a binary include/exclude tree and
searching it with MCTS.  The resulting initial set of features will
likely contain noisy along with useful features.  They use a filtering
technique to rank the best features in the set.  This differs from our
approach, as we measure MI for the entire feature set without using
filtering methods that can potentially discard correlations among
features.

\end{section}

\begin{section}{Results~\label{sec:results}}
We compare MI measurements from standard KSG and from our tree-search
modification of it (T-KSG) as the number of dimensions and the amount of
noise increases. For each of the two tests below, we begin by making a
baseline MI measurement using standard KSG~\footnote{We verify the
baseline MI value by processing the data through an ML algorithm,
measuring MI on the output ($I_f$) using the histogram method, and
comparing it to that baseline KSG measurement.  I.e., we use the fact
that $I_f$ is trivial to compute and then confirm that $I_i = I_f$ as
discussed in section~\ref{subsec:seplimit}.~\label{foot:verifyMI}}.  We
then begin adding noise and increasing the dimension of the variable
space. We want the new noise variables to be similar to the existing
data because added noise with, for example, a mean that is very
different from the existing variables, would be easier to identify as
noise. Thus, to create each noise variable, we randomly choose one of
the existing variables, shuffle it, and then include the shuffled values
back into in the data as a new variable. We repeat this process for each
noise variable we add.  The result is a dataset that has the original
variables, which each contain some information about the class of each
event, and some number of additional variables that give no information
about event class.  We can then compare the MI measurements from
standard KSG to those from T-KSG in these noisy datasets.

\begin{subsection}{Spherical Gaussian Test~\label{subsec:gauss}}

For the first test, we use partially overlapping Gaussian distributions
for signal and background and then add increasing amounts of noise.  The
signal data is a 5-dimensional Gaussian distribution with its center at
$(0,0,0,0,0)$ and a diagonal covariance matrix with 1's along the
diagonal.  The background data have the same covariance matrix but are
shifted to have their center at $(1,1,1,1,1)$. KSG's estimate of the MI
for the data without any added noise is $\langle$MI$\rangle =
0.566\pm0.002$ bits. As discussed in footnote~\ref{foot:verifyMI}, we
verified the accuracy of this baseline MI value.

We then added 6, 12, 25, 50, and finally 100 noise variables to the
data, and the results are shown in Figure~\ref{fig:gaussTest} and
Table~\ref{tab:gaussTest}.  Column one of Table~\ref{tab:gaussTest}
shows the amount of noise added. Column two shows the widely known
problem that KSG rapidly begins to underestimate MI as noise and
dimension increases. Column three, which is the key result, shows that
T-KSG is able to maintain an accurate estimate of MI even as noise and
the dimension of the dataset grows and becomes increasingly
sparse~\footnote{Because the added noise variables contain no
information about the event classes, and are uncorrelated with any other
variables, they each decrease the density of the variable space.}.
Column four shows the number of noise variables that T-KSG included in
its final/optimum path (i.e., its chosen set of variables), and it shows
why T-KSG is so effective. Adding noise variables to KSG's input
decreases its estimate of MI, and so as the T-KSG algorithm follows its
procedure for finding the set of variables that maximizes MI, it will
favor those paths that have fewer noise variables. Our T-KSG method
could reasonably be viewed as simply a method to avoid passing noise
into KSG.

Naively, one might expect T-KSG to reject all noise variables, which
would correspond to all zeros in column four.  That it can't is
understandable by recognizing that for any optimization algorithm that
calculates its FOM based on random sub-samples from a dataset, the FOM
will have some statistical fluctuations even when it is at the optimum
point.  And so at that optimum point, the algorithm will be indifferent
among options that are within that width.  Looking at the first two
entries in column two, one sees that, for this dataset, including as
many as six noise variables reduces KSG's estimate of MI by only about
2\%.  Because this is within the range of fluctuations among different
sub-samples, T-KSG is indifferent among paths that are within a few
percent of one another.

Over many runs and a wide range of noise levels and dimensions, we found
that T-KSG's optimum path consistently included all five good variables,
excluded all but a few noise variables, and maintained an extremely
accurate estimate of MI.  This held even in the most extreme case of
adding 100 noise variables to the five original variables -- a space
that would obviously be intractable to search via brute force.

\begin{table}[!htb]
  \centering
    \begin{tabular}{|c|c|c|c|}
    \hline
    \# Noise & \makecell{Standard\\KSG [bits]} & T-KSG [bits] & \makecell{Noise\\ Included}\\ 
    \hline
    \hline
     0 & $0.566(2)$ & $0.566(2)$ & n/a\\ \hline
     6 & $0.556(4)$ & $0.567(1)$ & $0-2$\\ \hline
     12 & $0.520(5)$ & $0.563(2)$ & $2-4$\\ \hline
     25 & $0.442(3)$ & $0.564(2)$ & $4-5$\\ \hline
     50 & $0.340(2)$ & $0.570(1)$ & $5-7$\\ \hline
     100 & $0.228(3)$ & $0.561(2)$ & $3-5$\\
    \hline
  \end{tabular}
  \caption{{\bf Gaussian Test.} MI calculations from standard KSG and our
  T-KSG method for a Gaussian dataset with a true MI of $0.566\pm0.002$ bits in
  the presence of varying amounts of noise. Column~1 shows the number of noise
  variables added to the data, column~2 shows that the MI estimate from
  standard KSG falls rapidly with increasing noise, column~3 shows the our
  estimate of MI is unaffected even by large amounts of noise, and column~4
  shows how many of the noise variables were typically included by our
  algorithm. T-KSG typically took a 5-60 minutes for the tests with 0-50 added
  noise dimensions, and several hours for the test with 100 added noise
  dimensions.  Details on the machine used for testing are given
  elsewhere~\ref{foot:specs}.\label{tab:gaussTest}}
\end{table}

\begin{figure}[!htb]
\centering
\includegraphics[width=3.6in]{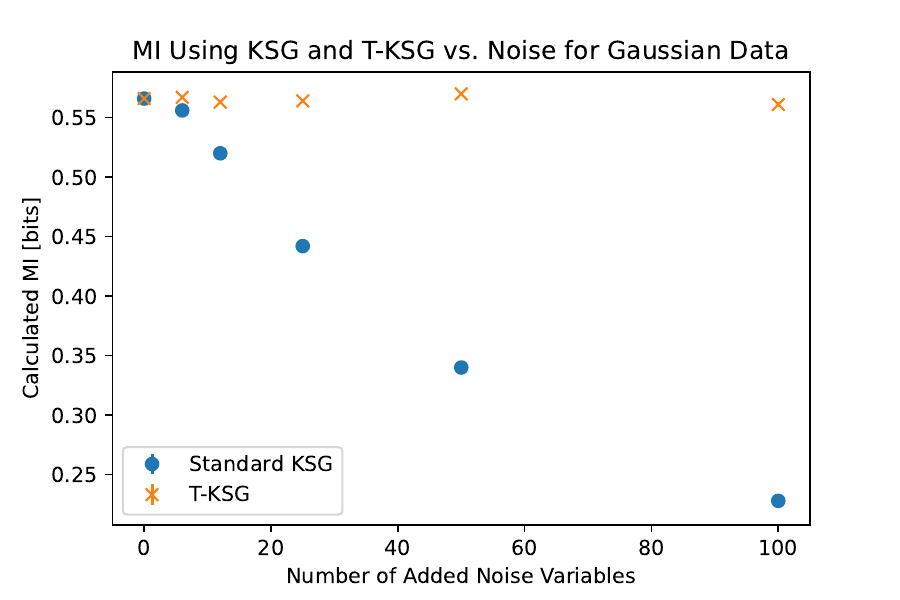}
\hfill
  \caption{{\bf Gaussian Test.} MI calculations from standard KSG and our
  T-KSG method for a Gaussian dataset with a true MI of $0.566\pm0.002$
  bits in the presence of varying amounts of noise.   This is the same
  data shown in Table~\ref{tab:gaussTest}.  One sees clearly that our
  T-KSG estimate of MI remains stable and accurate even as noise and
  sparsity increase, while standard KSG begins to significantly
  underestimate MI as noise increases.~\label{fig:gaussTest}}
\end{figure}

\end{subsection}

\begin{subsection}{SUSY Test~\label{subsec:susy}}

The Gaussian test of T-KSG in section~\ref{subsec:gauss} worked well,
but we would like to test it on more realistic data.  Here, we test
T-KSG on a signal-background pair of data samples produced by Baldi,
Sadowski, and Whiteson (BSW)~\cite{whiteson2014deeplearning} for their
study on the use of deep-learning in a search for supersymmetry (SUSY)
on the ATLAS experiment~\cite{atlas}. The signal data simulates the
production of the supersymmetric $\chi^\pm$ which decays to $W$ bosons
and a supersymmetric $\chi^0$. The $W$'s then decay to charged leptons
and neutrinos. Although the $\chi^0$ and the neutrinos are not directly
observable, their presence can be inferred from momentum imbalance and
missing energy in the detector. The background events are a Standard
Model process in which a pair of $W$ bosons are produced and then each
decay to a charged lepton and a neutrino, which is a commonly produced
event at ATLAS with a signature that is similar to the signal events.
BSW produced the event samples using Madgraph~\cite{madgraph2011} and
Pythia~\cite{pythia2006}, and used Delphes~\cite{delphes2009} to
simulate the response of the ATLAS detector. The result is a signal
dataset and a background dataset with 18 kinematic variables that mimic
what one would see when searching the ATLAS data for production of this
SUSY decay mode~\footnote{Their datasets are publicly
available~\cite{uci.ml.repo}}.

The details of the 18 kinematic events are not needed here, and are
described elsewhere~\cite{whiteson2014deeplearning}. Briefly, there are
eight variables that describe the momentum and angle of the two leptons
along with the energy and momentum imbalance of the event. There are
then ten additional variables that are functions of the first eight.

Searching for SUSY in the real data is a standard binary classification
problem in which one first uses the distributions of the 18 variables in
the simulated data to develop a model that distinguishes between the two
event types and then applies that model to the real data.  In practice,
this often involves using an ML technique to create the model from the
simulated data that discriminates between the event classes using the
features in the signal and background datasets.

Our test of T-KSG on the SUSY data is functionally identical to the
Gaussian test.  We have two datasets which partially overlap and we want
to estimate MI as a measure of how distinguishable, in principle, the
signal is from the background. One difference between the two tests is
that while every variable in the Gaussian study provides independent
information about the class of an event, the SUSY data has, as described
above, \emph{primitive} variables (the first eight) and \emph{derived}
variables (the additional ten).  Because the derived variables are
simply functions of the primitive variables, they add no new information
about which sample a given event is from.  In a previous
work~\cite{Carrara2017upperlimit}, we found, as expected, that after
calculating the amount of information that the primitive variables
contained about event type, the derived variables did not add any new
information.  I.e., they did not increase KSG's estimate of
MI~\footnote{In~\cite{Carrara2017upperlimit} we actually used the
Jensen-Shannon divergence between the signal and background samples as
our metric rather than mutual information between the input data and the
answer. However, for binary a classification problem the two quantities
are the same.}.

We will follow the same procedure to compare KSG with T-KSG that we used
in the Gaussian study:  We initially estimate MI using KSG and T-KSG on
the dataset without adding any noise, and then add increasing amounts of
noise to compare the stability of T-KSG to that of KSG.

Without noise, KSG and T-KSG give a consistent result of
($0.362\pm0.003$~bits) as each other, and the same as the result found
in Table~3 of~\cite{Carrara2017upperlimit}~\footnote{As in the Gaussian
study, we verified this baseline value by processing the data through a
neural network and measuring the MI of the one-dimensional output using
a simple histogram method.}.  The estimates from KSG and T-KSG as noise
is added are shown in Table~\ref{tab:SUSYTest} and
Figure~\ref{fig:SUSYTest}.  As in the Gaussian test, we see that as
noise increases, KSG's underestimate of MI worsens, while T-KSG's
estimate remains stable.

Column three of Table~\ref{tab:SUSYTest} shows that T-KSG typically
includes $\sim2$ noise variables.  Although these don't alter T-KSG's
estimate of MI, it is interesting to note that fewer noise variables are
included in this test ($\sim2$) than were typically included in the
Gaussian test ($\sim4$).  As discussed in section~\ref{subsec:gauss},
T-KSG is insensitive to additions of \emph{small} numbers of noise
variables, where \emph{small} means few enough that KSG's estimate of MI
does not shift outside of statistical fluctuations and so doesn't alter
T-KSG's choice of an optimum variable set.  Comparing column~2 in
Tables~\ref{tab:gaussTest} and \ref{tab:SUSYTest} shows that for the
Gaussian data, standard KSG underestimates MI by only about $2\%$ when 6
noise variables are added, while for the SUSY data, with the same amount
of noise, the underestimate is $19\%$.  This makes T-KSG more sensitive
to added noise in the SUSY case and hence it will include fewer noise
variables in its optimum set.

\begin{table}[!htb]
  \centering
    \begin{tabular}{|c|c|c|c|}
    \hline
    \# Noise & \makecell{Standard\\KSG [bits]} & T-KSG [bits] & \makecell{Noise\\ Included}\\ 
    \hline
    \hline
     0 & $0.359(2)$ & $0.365(2)$ & n/a\\ \hline
     6 & $0.293(2)$ & $0.367(2)$ & $0-1$\\ \hline
     12 & $0.250(2)$ & $0.365(2)$ & $0-2$\\ \hline
     25 & $0.191(2)$ & $0.359(2)$ & $1-2$\\ \hline
     50 & $0.115(2)$ & $0.349(2)$ & $1-2$\\ \hline
     100 & $0.069(1)$ & $0.356(2)$ & $1-4$\\
    \hline
  \end{tabular}
  \caption{{\bf SUSY Test.} MI calculations from standard KSG and our
  T-KSG method for a SUSY dataset with a true MI of $0.362\pm0.003$ bits in the
  presence of varying amounts of noise. Column~1 shows the number of noise
  variables added to the data, column~2 shows that the MI estimate from
  standard KSG falls rapidly with increasing noise, column~3 shows the our
  estimate of MI is unaffected even by large amounts of noise, and column~4
  shows how many of the noise variables were typically included by our
  algorithm. T-KSG typically took a 5-60 minutes for the tests with 0-50 added
  noise dimensions, and several hours for the test with 100 added noise
  dimensions.  Details on the machine used for testing are given
  elsewhere~\ref{foot:specs}.\label{tab:SUSYTest}}
\end{table}

\begin{figure}[!htb]
\centering
\includegraphics[width=3.4in]{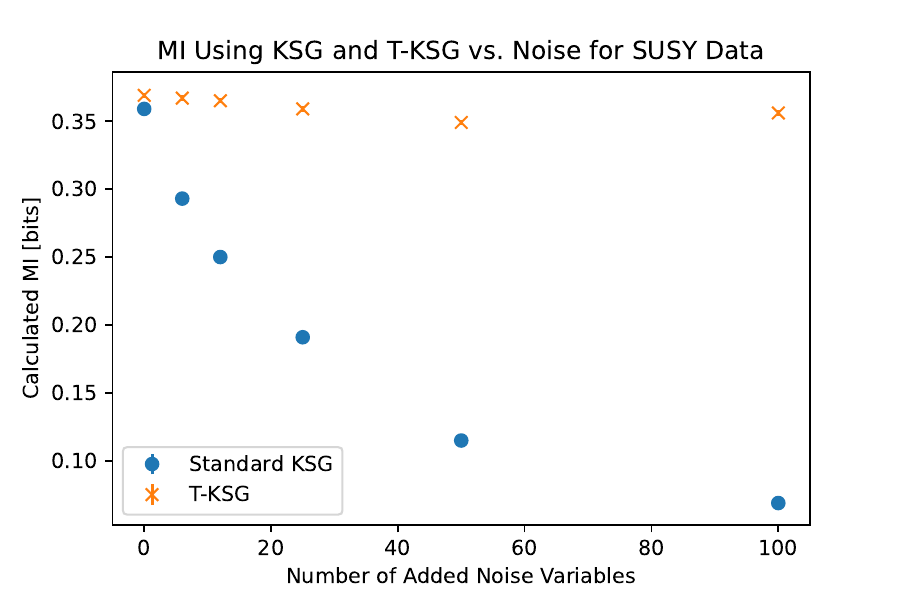}
\hfill
  \caption{{\bf SUSY Test.} MI calculations from standard KSG and our
  T-KSG method for a SUSY dataset with a true MI of $0.362\pm0.003$ bits
  in the presence of varying amounts of noise.   This is the same data
  shown in Table~\ref{tab:SUSYTest}.  One sees clearly that our T-KSG
  estimate of MI remains stable and accurate even as noise and sparsity
  increase, while standard KSG begins to significantly underestimate MI
  as noise increases.~\label{fig:SUSYTest}}
\end{figure}

\end{subsection}

\end{section}

\begin{section}{Conclusion~\label{sec:conclusion}}

MI is a centrally important quantitative measure of the correlations among
variables.  It is widely used in data analysis directly, and in studying models
produced by ML algorithms.  Despite its importance, existing methods for
computing MI perform poorly in higher dimensions or in the presence of noise.

We presented our T-KSG method for accurately estimating MI even in higher
dimensions and in the presence of noisy variables.  T-KSG is a combination of
the standard KSG method for measuring MI and the MCTS method that was developed
to improve computers' ability to play turn-based strategy games.  To improve
KSG, we developed a modified MCTS method that allows us to use a continuous
result and to rapidly remove variables that would cause standard KSG to
underestimate MI.

We have tested our method on a five-dimensional Gaussian sample and on
an 18-dimensional sample of simulated SUSY data.  For both studies, we
verified that in the absence of noise, T-KSG and standard KSG give the
same result as each other and the same result as an independent estimate
of MI. We then added between 6 and 100 extra dimensions of noise and
showed that T-KSG's estimate of MI was stable while KSG's estimate fell
by $40-80\%$.  Our software implementation of T-KSG is publicly
available on \href{https://github.com/jesse794/T-KSG}{GitHub}.

In addition to introducing the T-KSG method, we also reiterated
discussions from a previous work on the importance of MI as: 1) an
absolute measure of separability, 2) an absolute measure of an ML
model's performance, and 3) a way to rapidly carry out feature selection
in high-dimensional data.

For convenience, we gather the most used terms and acronyms in
Table~\ref{tab:glossary}.

\begin{table}[!htb]
  \centering
  \begin{tabularx}{\linewidth}{>{\hsize=0.35\hsize}X>{\hsize=1.65\hsize\arraybackslash}X}
    \hline
    Term & Meaning\\
    \hline

     \dtmi & Our term for the absolute amount of information lost by a
     transformation (often an ML model). \\

     DPI & Data Processing Inequality~~\cite{coverthomas} \\

     $I_f$ & Our term for the MI between the output of a transformation
     (often an ML model) and the true classes of the events.  \\

     $I_i$ &  Our term for the MI between descriptive variables and the
     true classes of the events. \\

     KL & An approach developed by L.F.\ Kozachenko and N.N.\ Leonenko
     for estimating the Shannon entropy of a continuous
     variable~\cite{KL}. \\

     KSG & The most widely used method for computing MI.\ Developed by
     Kraskov, Stögbauer, and Grassberger~\cite{ksg2004mi}. \\

     MCTS & The Monte Carlo Tree Search method. \\

     MI & Mutual Information \\

     ML & Machine Learning \\

     T-KSG & Our approach to computing MI, based on a combination of the
     KSG and MCTS methods. \\

    \hline
  \end{tabularx}
    \caption{Glossary of the most used terms and acronyms.\label{tab:glossary}}
  \end{table}

\end{section}

\begin{section}{Acknowledgments}

We thank Ariel Caticha and Greg Ver Steeg for helpful discussions. N.C.\
thanks the \textit{Nuclear Science and Security Consortium} for research support.

This material is based upon work supported by the Department of Energy National Nuclear Security Administration through the Nuclear Science and Security Consortium under Award Number DE-NA0003996.

\paragraph{Disclaimer}
This report was prepared as an account of work sponsored by an agency of the United States Government.  Neither the United States Government nor any agency thereof, nor any of their employees, makes any warranty, express of limited, or assumes any legal liability or responsibility for the accuracy, completeness, or usefulness of any information, apparatus, product, or process disclosed, or represents that its use would not infringe privately owned rights.  Reference herein to any specific commercial product, process, or service by trade name, trademark, manufacturer, or otherwise does not necessarily constitute or imply its endorsement, recommendation, or favoring by the United States Government or any agency thereof.  The view and opinions of authors expressed herein do not necessarily state or reflect those of the United States Government or any agency thereof.
\end{section}

\bibliographystyle{unsrt}
\bibliography{refs}

\end{document}